\documentclass{article}
\usepackage{spconf,amsmath,epsfig}

\newcommand{\Octave}{\textit{Octave}}
\newcommand{\Matlab}{\textit{Matlab}}
\newcommand{\ifft}{\textit{IFFT}}
\newcommand{\ofdm}{\textit{OFDM}}

\title{A Fixed-Point Type for Octave}

\name{David Bateman, Laurent Mazet, V\'eronique Buzenac-Settineri and Markus Muck}
\address{Motorola Labs Paris \\
        Parc Les Algorithmes, Saint Aubin \\
	91193 Gif-Sur-Yvette Cedex - France \\ \\
        David.Bateman@motorola.com, Laurent.Mazet@motorola.com, \\
	Veronique.Buzenac@motorola.com, Markus.Muck@motorola.com}

\begin{document}

\maketitle

\begin{abstract}
This paper announces the availability of a fixed point toolbox for the 
{\Matlab} compatible software package {\Octave}. This toolbox is 
released under the GNU Public License, and can be used to model the losses
in algorithms implemented in hardware. Furthermore, this paper presents
as an example of the use of this toolbox, the effects of a fixed point
implementation on the precision of an {\ofdm}  modulator.
\end{abstract}

\section{Introduction}
\label{sec:intro}

When implementing algorithms in hardware, it is common practice to
reduce the accuracy of the representation of numbers to a smaller
number of bits. This allows much lower complexity in the hardware, at
the cost of accuracy and potential overflow problems. Such
representations are known as fixed point~\cite{iscs.sung91}.  

Many previous authors have presented solutions for mo\-del\-ling fixed
point representations~\cite{vlsi.kim95,date.keding98,scp.harton99}. A
common point of all of these solution is that they are written in a
low-level programming language such as {\it C} or {\it C++}. In
addition the only code that is publically available is that presented
by Kim et al~\cite{vlsi.kim95}, and it is released under a license
limiting its use to academic use only.

However, being low level implementations of the fixed point
representations, they don't allow the engineer developing an algorithm
the freedom to easily test multiple ideas and their consquence on
their fixed point implementations. This gap between an algorithms
development and its implementation can result in implementations that
are overly complex and/or the choice of the algorithm to implement
being sub-optimal.

For this reason it is important to consider the implementation losses
in algorithms, due to fixed point representations, early in their
design. To do this, there is a clear need for support of the analysis
of fixed point types in standard high-level engineering design
tools. One such tool that is used by most engineers is
{\Matlab}~\cite{matlab.website} or its open-source cousin
{\Octave}~\cite{octave.website}. {\Matlab} supports fixed point types
through its {\it SimulLink} package~\cite{matlab.fixedpoint}.

However, the authors of this paper have preferred to write a toolbox
implementation of a fixed point type to ease the softwares using for
those not having {\Matlab} licenses, but equally for issues of the
softwares use on parallel computers. The authors have equally made
this toolbox available under the GNU Public License as part of the
{\it Octave-Forge} package~\cite{octaveforge.website}. To the authors
knowledge this is the first time a fixed point toolkit for a high
level engineering tool is available that intrinsically treats real,
complex and matrix fixed point types. It is equally the first time
that such code is publicly available that allows its use for the
development of algorithms in commerical applications, with the sole
restriction that changes to the fixed point type itself are returned
to the community.

This article, describes the contents and use of the fixed point
toolbox, giving useful examples and limitations. In addition, as an
example of the use of this toolbox, we analyses the effects of a
fixed-point implementation on an {\ofdm} modulator.

\section{Description of the Code}

\subsection{Representation of Fixed Point Numbers}

Fixed point numbers can be represented digitally in several
manners, including {\it sign-magnitude}, {\it ones-complement} and
{\it twos-complement}.  However, the {\it twos-complement}
representation simplifies the implementation of many operators in
hardware, and so it is most common to see the {\it twos-complment}
representation uses. This toolbox therefore represents fixed
point objects using the {\it twos-complement} representation. All
fixed point objects in this toolbox are represented by a {\it
long int} that is used in the following manner

\begin{itemize}
\item 1 bit representing the sign,
\item $is$ bits representing the integer part of the number, and
\item $ds$ bits representing the decimal part of the number.
\end{itemize}

The numbers that can then be represented are then given by

\begin{equation}
- 2 ^ {is} \leq x \leq 2 ^ {is} - 1
\end{equation}

and the distance between two values of $x$ that are not represented by
the same fixed point value is $2^{-ds}$.

The number of bits that can be used in the representation of the
fixed point objects is determined by the number of bits in a {\it
long int} on the platform. Valid values include 32- and
64-bits. However, to simplify their lives, the authors have also
chosen to reduce the available number of bits by 1, so that issues
with the differences between representations of {\it long int} and
{\it unsigned long int} need not be taken into account. Therefore
valid values of {\it is} and {\it ds} are given by

\begin{equation}
0 \leq \left( is + ds \right) \leq n - 2 
\end{equation}

where $n$ is either 32 or 64, depending on the number of bits in a
{\it long int}. It should be noted that given the above criteria it is
possible to create a fixed point representation that lacks a
representation of the number 1. This makes the implementation of
certain operators difficult, and so the valid representations are
further limited to

\begin{equation}
0 \leq \left( is , ds , is + ds \right) \leq n - 2 
\end{equation}

This does not mean that other numbers can not be represented by this toolbox,
but rather that the numbers must be scaled prior to their being represented.

This toolbox allows both fixed point real and complex scalars to
be represented, as well as fixed point real and complex matrices. 
The real and imaginary parts of the fixed point numbers and each element 
of fixed point matrices having their own fixed point representation.

\subsection{Creation of Fixed Point Numbers}

Before using a fixed point type, some variables must be created that use this
type. This is done with the function {\it fixed}. The function {\it fixed}
can be used in several manners, depending on the number and type of the
arguments that are given. It can be used to create scalar, complex, matrix
and complex matrix values of the fixed type.

The generic call to {\it fixed} is {\it fixed( is, ds, f)},
where the variables {\it is}, {\it ds} are as previously described. The
variable {\it f} can be either a scalar, complex, matrix or complex matrix 
of values that will be converted to a fixed point representation. It can
equally be another fixed point value, in which case {\it fixed} has the
effect of changing the representation of {\it f} to another representation
given by {\it is} and {\it ds}.

If matrices are used for {\it is}, {\it ds}, or {\it f}, then the dimensions
of all of the matrices must match. However, it is valid to have {\it is} or
{\it ds} as scalar values, which will be expanded to the same dimension 
as the other matrices, before use in the conversion to a fixed point value.
The variable {\it f} however, must be a matrix if either {\it is} or {\it ds}
is a matrix.

The most basic use of the function {\it fixed} can be seen in the example

{\footnotesize
\begin{verbatim}
octave:1> a = fixed(7, 2, 1)
ans = 1
octave:2> isfixed(a)
ans = 1 
octave:3> typeinfo(a)
ans = fixed scalar
\end{verbatim}
}

which demonstrates the creation of a real scalar fixed point value
with 7 bits of precision in the integer part, 2 bits in the decimal
part and the value 1. The function {\it isfixed} can be used to
identify whether a variable is of the fixed point type or
not. Equally, using the {\it typeinfo} or {\it whos} function allows
the variable to be identified as "fixed scalar".

Other examples of valid uses of {\it fixed} are

{\footnotesize
\begin{verbatim}
octave:1> a = fixed(7, 2, 1);
octave:2> b = fixed(7, 2+1i, 1+1i);
octave:3> c = fixed(7, 2, 255*rand(10, 10) - 128);
octave:4> is = 3 * ones(10,10) + 4*eye(10); 
octave:5> d = fixed(is, 1, eye(10));
octave:6> e = fixed(7, 2, 255*rand(10, 10) -
>   128 + 1i*(255 * rand(10, 10) - 128));
\end{verbatim}
}

With two arguments given to {\it fixed}, it is assumed that {\it f} is
zero or a matrix of zeros, and so {\it fixed} called with two
arguments is equivalent to calling with three arguments with the third
arguments being zero. For example

{\footnotesize
\begin{verbatim}
octave:1> a = fixed([7, 7], [2, 2], zeros(1,2));
octave:2> b = fixed([7, 7], [2, 2]);
octave:3> assert(a == b);
\end{verbatim}
}

Called with a single argument {\it fixed}, and a fixed point argument,
{\it b = fixed(a)} is equivalent to {\it b = a}. If {\it a} is 
not itself fixed point, then the integer part of {\it a} is used to create
a fixed point value, with the minimum number of bits needed to represent
it. For example

{\footnotesize
\begin{verbatim}
octave:1> b = fixed(1:4);
\end{verbatim}
}

creates a fixed point row vector with 4 values. Each of these values
has the minimum number of bits needed to represent it. That is {\it  b(1)}
uses 1 bit to represent the integer part, {\it b(2:3)} use 2 bits and 
{\it b(4)} uses 3 bits. The single argument used with {\it fixed} can 
equally be a complex value, in which case the real and imaginary parts are
treated separately to create a composite fixed point value.

\subsection{Overflow Behavior of the Fixed Point Type}

When converting a floating point number to a fixed point number the
overflow behavior of the fixed point type is such that it implements
clipping of the data to the maximum or minimum value that is
representable in the fixed point type. This effectively simulates the
behavior of an analog to digital conversion. For example

{\footnotesize
\begin{verbatim}
octave:1> a = fixed(7, 2, 200)
a = 127.75
\end{verbatim}
}

However, the overflow behavior of the fixed point type is distinctly
different if the overflow occurs within a fixed point operation itself.
In this case the excess bits generated by the overflow are dropped.
For example

{\footnotesize
\begin{verbatim}
octave:1> a = fixed(7, 2, 127) + fixed(7, 2, 2)
a = -127
octave:2> a = fixed(7, 2, -127) + fixed(7, 2, -2)
a = 127
\end{verbatim}
}

The case where the representation of the fixed point object changes is
different again. In this case the sign is maintained, while the
most-significant bits of the representation are dropped. For example

{\footnotesize
\begin{verbatim}
octave:1> a = fixed(6, 2, fixed(7, 2, -127.25))
a = -63.25
octave:2> a = fixed(6, 2, fixed(7, 2, 127.25))
a = 63.25
octave:3> a = fixed(7, 1, fixed(7, 2, -127.25))
a = -127.5
octave:4> a = fixed(7, 1, fixed(7, 2, 127.25))
a = 127
\end{verbatim}
}

In addition to the overflow issue discussed above, it is important to
take into account what happens when and operator is used on two fixed 
point values with different representations. For example

{\footnotesize
\begin{verbatim}
octave:1> a = fixed(7, 2, 1);
octave:2> b = fixed(6, 3, 1);
octave:3> c = a + b;
octave:4> d = [c.int, c.dec]
d =
  7  3
\end{verbatim}
}

as can be seen the representation of the output fixed point value is 
promoted such that {\it c.int = max(a.int,b.int)}
and {\it c.dec = max(a.dec,b.dec)}. If this promotion causes the 
maximum number of bits in a fixed point representation to be exceeded,
then an error will occur.

\subsection{Analysis of Complexity}

After the fixed point type is first used, four variables are
initialized.  The {\it fixed\_point\_count\_operations} variable is of
particular interest.  The {\Octave} fixed point type can keep track
of all of the fixed point operations and their type. This is very
useful for a simple complexity analysis of the algorithms. To allow
the fixed point type to track operations the variable {\it
fixed\_point\_count\_operations} must be non-zero. The function
{\it reset\_fixed\_operations}, can be used to reset the number of
operations since the last reset as given by the 
{\it display\_fixed\_operations} function.

\subsection{Accessing Internal Fields}

Once a variable has been defined as a fixed point object, the parameters of the
field of this structure can be obtained by adding a suffix to the variable.
Valid suffixes are '.x', '.sign', '.int' and '.dec', which return

\begin{itemize}
\item[{\it .x}]
The floating point representation of the fixed point number
\item[{\it .sign}]
The sign of the fixed point number
\item[{\it .int}]
The number of bits representing the integer part of the fixed point number
\item[{\it .dec}]
The number of bits representing the decimal part of the fixed point number
\end{itemize}

As each fixed point value in a matrix can have a different number of bits 
in its representation, these suffixes return objects of the same size as
the original fixed point object. For example

{\footnotesize
\begin{verbatim}
octave:1> a = [-3:3];
octave:2> b = fixed(7, 2, a);
octave:3> b.sign
ans =
  -1  -1  -1   0   1   1   1
octave:4> b.int
ans =
  7  7  7  7  7  7  7
octave:5> b.dec
ans =
  2  2  2  2  2  2  2
octave:5> c = b.x;
octave:6> typeinfo(c)
ans = matrix
\end{verbatim}
}

The suffixes {\it .int} and {\it .dec} can also be used to change the internal
representation of a fixed point value. This can result in a loss of
precision in the representation of the fixed point value, which models
the same process as occurs in hardware. For example

{\footnotesize
\begin{verbatim}
octave:1> b = fixed(7, 2, [3.25, 3.25]);
octave:2> b(1).dec = [0, 2];
b =
     3  3.25
\end{verbatim}
}

However, the value itself should not be changed using the suffix {\it .x}. 

\subsection{Function Overloading}

An important consideration in the use of the fixed point toolbox is
that many of the internal functions of {\Octave}, such as {\it
diag}, can not accept fixed point objects as an input. This package
therefore uses the {\it dispatch} function of {\it Octave-Forge} to
overload the internal {\Octave} functions with equivalent
functions that work with fixed point objects, so that the standard
function names can be used. However, at any time the fixed point
specific version of the function can be used by explicitly calling its
function name. There are too many functions available for use with the
fixed point type to list in this article, and so interested readers are
referred to the software package itself~\cite{octaveforge.website}

\subsection{Putting it all Together}

Now that the basic functioning of the fixed point type has been discussed,
it is time to consider how to put all of it together. The main advantage of 
this fixed point type over an implementation of specific fixed point code, 
is the ability to define a function once and use as either fixed or floating 
point. Consider the example

{\footnotesize
\begin{verbatim}
function [b, bf] = testfixed(is, ds, n)
a = randn(n, n);
af = fixed(is, ds, a);
b = myfunc(a, a);
bf = myfunc(af, af);
endfunction

function y = myfunc(a, b)
y = a + b;
endfunction
\end{verbatim}
}

In this case {\it b} and {\it bf} will be returned from the function
{\it testfixed} as floating and fixed point types respectively, while
the underlying function {\it myfunc} does not explicitly define that it
uses a fixed point type. This is a major advantage, as it is critical to
understand the loss of precision in an algorithm when converting from
floating to fixed point types for an optimal hardware implementation.
This mixing of functions that treat both floating and fixed point types
can even apply to Oct-files (The {\Octave} equivalent of a {\Matlab}
mex-file), as will be discussed later.

The main limitation to the above is the use of the concatenation operator,
such as {\it [a,b]}, that is hard-coded into current versions of {\Octave}
and is thus not aware of the fixed-point type. Therefore, such code should
be avoided and the function {\it concat} supplied with this package used
instead.

\section{The Fixed Point Type Applied to an {\ofdm} Modulator}

As an example of the use of the fixed point toolbox applied to a real
signal processing example, we consider the implementation of a Radix-4
{\ifft} in an {\ofdm} modulator~\cite{milcom.gifford01}. Code for this {\ifft} has been
written as a C++ template class, and integrated as an {\Octave} Oct-file.
This allowed a single version of the code to be instantiated to perform
both the fixed and floating point implementations of the same code. The
code for this example is available as part of the release of this 
software package.

A particular problem of a hardware implementation of an {\ifft} is that
each butterfly in the radix-4 {\ifft} consists of the summation of four
terms with a suitable phase. Thus, an additional 2 output bits are
potentially needed after each butterfly of the radix-4 {\ifft}. There are
several ways of addressing this issue

\begin{itemize}
\item Increase the number of bits in the fixed point representation by two 
 	after each radix-4 butterfly. There are then two ways of treating
	these added bits
	\begin{itemize}
	\item Accept them and let the size of the representation of the fixed
		point numbers grows. For large {\ifft}'s this is not acceptable
	\item Cut the least significant bits of representation, either after
		each butterfly, or after groups of butterflies. This reduces
		the number of bits in the representation, but still trades off
		complexity to avoid an overflow condition
	\end{itemize}
\item Keep the fixed point representation used in the {\ifft} fixed, but reduce
	the input signal level to avoid overflows. The {\ifft} can then have
	internal overflows.
\end{itemize}

An overflow will cause a bit-error which is not necessarily critical. The
last option is therefore attractive in that it allows the minimum complexity
in the hardware implementation of the {\ifft}. However, careful investigation of
the overflow effects are needed, which can be performed with the fixed point
toolbox discussed in this article.

The figure~\ref{fig:ofdm} below shows the case of a 64QAM {\ofdm} signal
similar to that used in the 802.11a standard. In this figure only the {\ofdm}
modulator has been represented using fixed point, while the rest of
the system is assumed to be perfect. Figure~\ref{fig:ofdm} shows the
tradeoff between the backoff of the RMS power in the frequency domain
signal relative to the fixed point representation for several
different fixed point representations.

\begin{figure}[htb]
  \begin{minipage}[b]{1.0\linewidth}
    \centering
    \centerline{\epsfig{figure=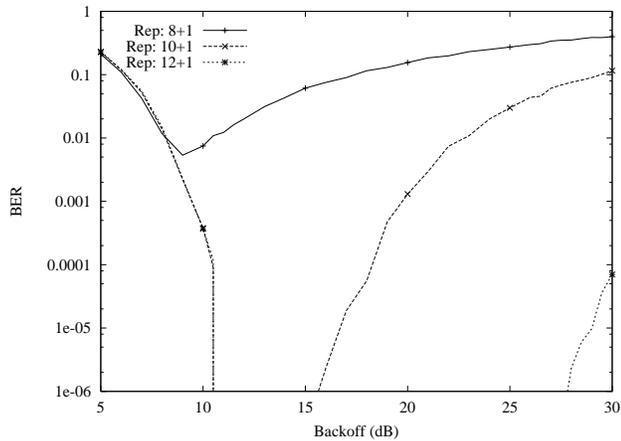,width=8.5cm}}
  \end{minipage}
  \begin{center}
  \caption{Bit-error rate due to fixed point representation for various 
	backoffs of the RMS power in frequency domain signal. Fixed point
	representation of {\it N} bits plus 1 bit for the sign}
  \label{fig:ofdm}
  \end{center}
\end{figure}

Two regions are clearly visible in figure~\ref{fig:ofdm}. When the
backoff of the RMS power is small, the effects of the overflow in the
{\ifft} dominate, and reduce the performance. When the backoff is large,
there are fewer bits in the fixed point representation relative to
the average signal power and therefore a slow degradation in the
performance. It is clear that somewhere between 11 and 13 bits in the
representation of the fixed point numbers in the {\ifft} is optimal,
with a backoff of approximately 13dB.

\section{Conclusion}

This article has announced the release of a public available package
for the analysis of fixed point implementations of algorithms within
{\Octave}. The code is available under the conditions of the GNU
Public License. The basic capabilities of this code has been discussed
and the simple examples of the code have been given.

Furthermore, this article has discussed the use of this package for 
the example of an {\ofdm} modulator using a particular radix-4 {\ifft}
implementation. The relationship between the clipping of the input
signal, the number of bits in the fixed point representation and the
noise introduced into the signal has been discussed.

\bibliographystyle{IEEEbib}
\bibliography{icassp2004}

\end{document}